\documentclass[preprint,preprintnumbers,amsmath,amssymb]{revtex4}

\usepackage{graphicx}% Include figure files
\usepackage{dcolumn}% Align table columns on decimal point
\usepackage{bm}% bold math

\begin{document}

\title{Fast convergence of white noise cross-correlation measurement archived by vector average in both frequency and time domain}

\author{Jian WEI}
\email{weijian@physics.rutgers.edu}
\affiliation{%
Department of Physics and Astronomy, Rutgers University,
Piscataway, NJ 08854
}%

\date{\today}% It is always \today, today,

\begin{abstract}
White noise measurement can provide very useful information in
addition to normal transport measurements. For example thermal
noise measurement can be used at sub Kelvin temperature to
determine the absolute electron temperature without applying any
heating current. And shot noise measurements helped to understand
the properties of nano and mesoscopic normal metal/superconductor
structures. But at low temperature and for relatively small
resistance it is difficult to measure the sample's noise magnitude
because the background thermal noise can be much larger and
usually there are other pick-up noises. Cross correlation
technique is one way to solve this problem. This article describes
an improved cross correlation algorithm that averages in both
frequency and time domain, and the realization of a simple
instrument set-up with PC and sound card. With this set-up it is
shown even with much larger background noise and pickup noises,
100pV/$\sqrt{Hz}$ white noise level can be easily measured in
seconds. Compared to the normally used cross-correlation methods,
it is several orders of magnitude faster.
\end{abstract}

\maketitle

\section{Introduction}

Recently a lot of attention has been paid on the properties of
S/N/S and N/N/N microbridge structures, and noise measurement
(shot noise \cite{Kum96, Reu03} as well as thermal noise
\cite{Hen99}) was used as an additional technique besides
transport measurements. But to measure the noise at cryogenic
temperature is difficult since the sample noise usually is much
smaller than the thermal noise of the components in the test
circuit. Noise thermometry for electrons at low temperature was
previously performed by current noise measurement with SQUID
\cite{Rou85}, which is very sensitive and has very low noise
level. However it is limited for small resistance and not for
voltage noise measurement, also it can not be used when magnetic
field is applied. Cross-correlation technique \cite{Sam99, Hen99,
Kum96} provides an alternative method to measure the sample noise
at low temperature. One bottleneck for this technique is that it
requires a lot of time to do cross-correlation to converge and
achieve the required sensitivity, and the trade off between
sensitivity and the time needed to converge makes it difficult to
use for low level noise experiments.

In this article we present an improved cross-correlation
algorithm as well as the test instrument set-up for measurement of
thermal noise. This algorithm does a vector average over both
time and specific frequency range.  It is worth noting that for
commercial spectrum analyzer usually only average over time is
used, and it is impossible to realize this algorithm within the
instrument because of limitations like memory and computation
speed. As shown in the following sections, much faster convergence
can be achieved with the new algorithm.

\section{Cross correlation principle}\label{cross}
\begin{figure}
\includegraphics{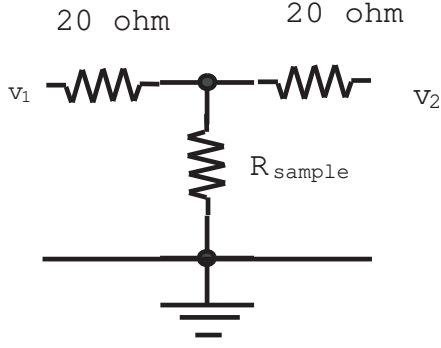}
\caption{\label{fig1} Schematic of the set-up. The sample
resistor is 10 $\Omega$ and 1.5$\Omega$. The thermal noise is 20
$\Omega$ resistor along the channel is much bigger than the
sample thermal noise.}
\end{figure}

It is well known 4-probe resistance measurement eliminates
contact resistance by measuring the voltage signal across the
sample that was stimulated by the current though the sample. It
can be also considered as measuring the "in phase" signal between
current and voltage across the sample. Cross-correlation is
similar in a sense it eliminates the channel's noise by measuring
the "similarity" or "in phase" signal between two different
voltage channels. To better illustrate it, let consider the
voltage signals from two channels. The Fourier components at
particular frequency $\omega$ are:
\begin{eqnarray}\label{eqn1}
    \vec{v_{1}}& = &Ae^{i\theta_{A}}+N_{1}e^{i\theta_{1}} \\
    \vec{v_{2}}& = &Ae^{i\theta_{A}}+N_{2}e^{i\theta_{2}}
\end{eqnarray}
Here A is the amplitude of the noise signal from our sample at
frequency $\omega$ and $N_{1}$, $N_{2}$ are the amplitude of
unwanted noise at $\omega$ generated in those two channels, for
example, the thermal noise generated by the 20$\Omega$ lead
resistor as shown in Fig.~\ref{fig1}.

To do cross-correlation we calculate the product of the two
vectors:
\begin{eqnarray}\label{eqn2}
    \vec{v} & = & \vec{v_{1}}\cdot \vec{v_{2}} \nonumber \\
    & = & A^{2}+N_{1}N_{2}e^{i(\theta_{2}-\theta_{1})}+AN_{1}e^{i(\theta_{A}-\theta_{1})}
    +AN_{2}e^{i(\theta_{2}-\theta_{A})}
\end{eqnarray}

For the last three terms in Eq.\ref{eqn2}, since all those phases
$\theta_{1}$, $\theta_{2}$ are random for white noise, and $A,
N_{1}$ and $N_{2}$ does not change over time, when we do the
average of $\vec{v}$ over enough long time, the random phase
terms will cancel each other and will give negligible
contribution to the total amplitude. So the real part of the
$\vec{v}$ will converge at $A^{2}$, and the imaginary part will
converge at $0$. Ideally if we do the average over infinite time
the amplitude will converge at $A^{2}$. But in practical
situation, the measurement time should be limited to some
reasonable extent.

To estimate the time needed to approach convergence limit, we need
to find when the deviation of the random phase term is much
smaller than the $A^{2}$ term. First in order to identify the
signals $\vec{v_{1}},\vec{v_{2}}$ at frequency $\omega$, the
sampling time $\Delta T$ should be much longer than $1/\omega$ to
get an accurate Fourier component. Then $\vec{v_{1}},\vec{v_{2}}$
at different times are acquired to do the vector average to
eliminate the random phase terms. If $N_{1}, N_{2} \gg A$, which
happens when the sample is at low temperature and the sample
signal amplitude $A$ is very small, it will take very long time to
approach the convergence limit, which requires large n in the
following equation:
\begin{eqnarray}\label{convergence}
N_{1}N_{2}\frac{\sum_{n}{e^{i(\theta_{2}-\theta_{1})}}}{n}\ll
A^{2} \nonumber \\
\overline{e^{i(\Delta\theta)}} =
\frac{\sum_{n}{e^{i(\theta_{2}-\theta_{1})}}}{n} \ll 1
  \end{eqnarray}

From standard textbook\cite{Yat99} we know for iid (independent
identical distributed) random sequence, n times average gives n
times smaller variance. Assuming our channel noises are iid, we
can expect that the variance of the averaged random phase term
$Var(\overline{e^{i(\Delta\theta)}})\propto 1/n$,  similarly for
the averaged correlation the variance
$Var(\overline{\vec{v}})\propto 1/n$, since our goal is to find
noise magnitude $A$, we need the variance of root square of real
part $\overline{\vec{v}}$, $Var(\sqrt{\overline{\Re{\vec{v}}}})$
which is proportional to $1/\sqrt{n}$, and finally what need
standard deviation to compare with $A$.  The standard deviation
$\sigma_{\sqrt{\overline{\Re{\vec{v}}}}}=\sqrt{Var(\sqrt{\overline{\Re{\vec{v}}}})}$
should be proportional to $n^{-1/4}$. The $-1/4$ exponent is
indeed observed in our experiment as shown in Sec.\ref{exponent}.
This $-1/4$ exponent might also be used as a criteria to decide
if the channel noise in different time steps can be fitted as iid
random sequence, i.e. whether there is some correlation in time
domain.

As indicated above by the $-1/4$ exponent, it is not very
effective to eliminate the channel noise by increasing the
measurement time steps. To accelerate the convergence process, a
new algorithm is described below. Consider cross-correlation
results at two different frequencies $\omega_m$ and $\omega_n$:
\begin{eqnarray}\label{eqn3}
    \vec{v_{m}}& = &A^{2}+N_{m}e^{i\theta_{m}} \\
    \vec{v_{n}}& = &A^{2}+N_{n}e^{i\theta_{n}}
\end{eqnarray}
here the last vector term $N_{m}e^{i\theta_{m}}$ represents the
vector sum of the last three terms in Eq.\ref{eqn2}. For "white"
noise, the amplitude $A$ is the same for different frequencies but
the phases $\theta_{m}$, $\theta_{n}$ are random. This means a
vector average over frequency domain is equivalent to the vector
average over time domain. Since in practice we usually acquire a
series of data points in one sampling time $\Delta T$ and then do
a FFT transform, we could compute all the points in the frequency
domain and use them to do vector average. For example the
commercial spectrum analyzer usually takes 1024 scaler voltage
points in $\Delta T$ (can't take more because of limited memory)
and give 512 vector points in the frequency domain. If we do
vector average of the 512 points, according to the above
statement it is similar to the result of average over 512 time
steps for one particular frequency, which means the convergence
at $A$ can be achieved 512 times faster. To test this we built
some simple experimental set-up as described in the following
section.

\section{Instrument set-up}

A schematic plot of the test set up is shown in Fig.~\ref{fig1}.
Since people usually use resistive stainless steel coax cable to
connect the sample in the low temperature stage, here a 20$\Omega$
resistor is used along each channel to simulate the channel
resistor. The sample resistors used here are 10$\Omega$ and
1.5$\Omega$ to simulate the low noise level from real sample. In
this case the amplitude of the sample noise $A$ is much smaller
than that of channel noise $N$, so it can only be retrieved by
cross-correlation technique.

The output of the two channels feeds separately to two PAR116
preamplifier (transformer mode) and PAR124 lock-in amplifier(only
as an additional cascade amplifier, the Monitor output is used).
The transformer is used to match the impedance. Two transformers
need to be similar because otherwise if may change the phase and
amplitude and affect the convergence. For example if there is a
fixed phase difference $\Delta\theta$ between two transformer, the
amplitude $A$ will be reduced to $cos(\Delta\theta)A$. After
amplification the signals are fed to left/right channel input of a
standard PCI sound card installed in a PII PC. The sound card then
digitize the signal with 16 bit resolution and 44.1kHz sampling
rates. After that the data is acquired by a program written in
Labview to the computer memory.  Then the program calculates
Fourier spectra and and performs correlation and vector average
etc., and shows all results on the screen in real time. A PC is
much better than a commercial analyzer when considering the memory
size and computation speed.

\section{Experimental results}\label{exponent}

\begin{figure}
\includegraphics{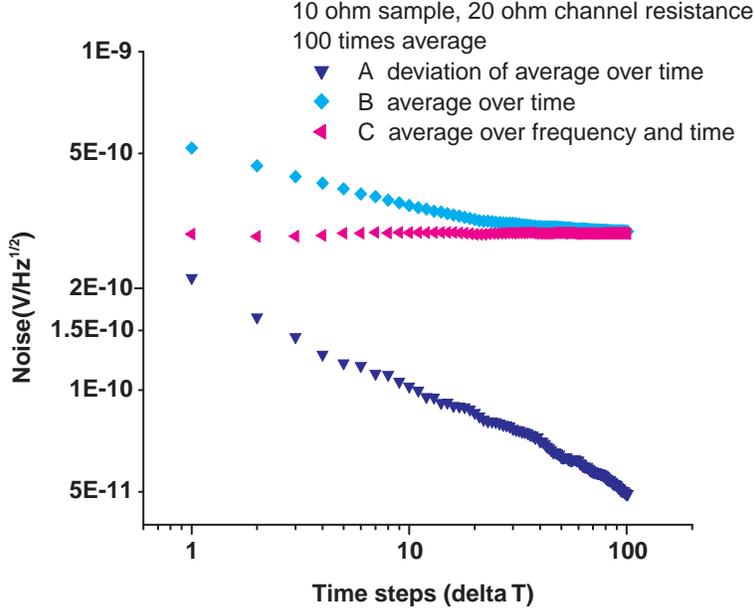}
\caption{\label{fig2} For a 10$\Omega$ sample resistor with
20$\Omega$ channel resistor, the cross-correlation results are
shown. Curve A shows standard deviation of the average over time.
The slope is proportional to $n^{-1/4}$. Curve B shows vector
average over time. Curve C shows vector average over both
frequency \emph{and} time domain. With this new algorithm the
convergence is achieved almost from the first point, much faster
than the conventional cross-correlation result shown by curve B.}
\end{figure}

In Fig.~\ref{fig2}, for the 10$\Omega$ sample resistor three
curves are shown to demonstrate the result of conventional
cross-correlation algorithm and to compare it with the new
algorithm. Curve A shows that the standard deviation of the
average over time, $\sigma_{\sqrt{\overline{\Re{\vec{v}}}}}$
decreases as the time elapse\cite{footnote}. As shown in
Sec.\ref{cross} it is proportional to $n^{-1/4}$, which can be
find easily from the log-log plot. Curve B shows the average over
time approaches slowly to $A$, which is around 0.3nV/$\sqrt{Hz}$,
after more than 100 times average. Curve C shows that when using
the new algorithm, the average over frequency \emph{and} time
almost converge at $A$ from the first point! In fact, curve A is
proportional to the difference between curve B and curve C. The
measured amplitude of $A$ is close to expected amplitude of $A$,
which is $\sqrt{4k_{B}TR}=$ 0.4nV/$\sqrt{Hz}$ for 10$\Omega$
sample resistor. This result is not bad when considering there
may be affects from non ideal phase and amplitude properties of
transformers and amplifiers, and uncertain pre-factors that came
in from the data processing like the use of windows when doing
FFT.

\begin{figure}
\includegraphics{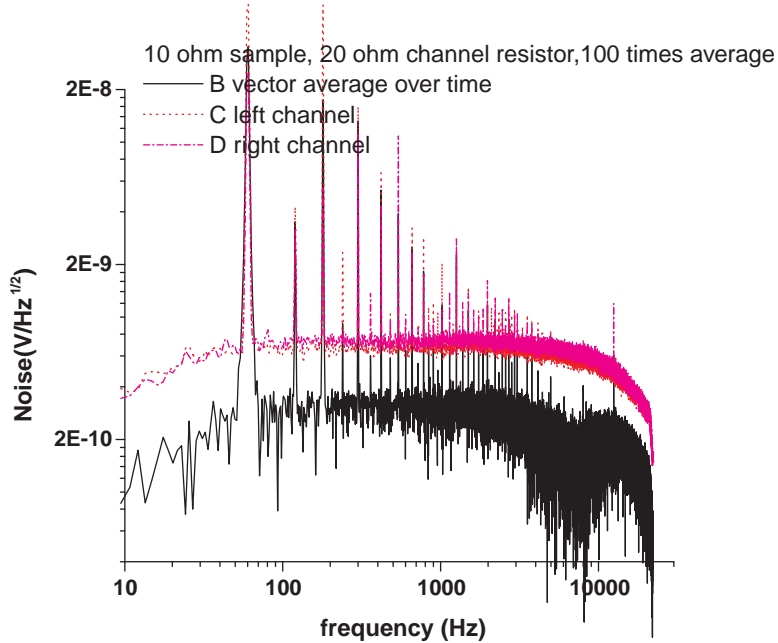}
\caption{\label{fig3} FFT spectrum after 100 times average. Curve
B shows the cross-correlation vector average over 100 time steps,
curve C, D shows the noise spectrum of left/right channel in
$\Delta T$. The noise floor level of curve B is close to
0.3nV/$\sqrt{Hz}$ as shown in Fig.~\ref{fig2}.  The decrease of
amplitude below 60Hz and the dip near 8kHz is due to the amplitude
and phase properties of transformers. The cut-off near 22kHz is
due to the Shannon limit, i.e., half of 44.1kHz sampling
frequency.}
\end{figure}

FFT spectrum after 100 times average is shown in Fig.~\ref{fig3}.
Curve B shows the result of the conventional cross-correlation
vector average over 100 time steps. For comparison, curves C, D
shows separately the noise spectrum of left/right channel measured
in $\Delta T$. The vector average over both frequency and time
domain is just a number, so it can not be shown in this frequency
spectrum figure. From curve B, despite of those pick up noise
peaks, we can still "see" the real noise level that is around
0.3nV/$\sqrt{Hz}$, which was also found by the program and shown
as the last point of curve B in Fig.~\ref{fig2}. The program
actually average \emph{scalarly} the spectrum amplitude from 1kHz
to 2kHz where the spectrum is almost flat and the affect of power
line noise peaks is smaller. And those points close to power line
noise peaks were abandoned. It is worth noting this
\emph{scalarly} average of amplitude over frequency is different
with the vector average over frequency.

With the presence of huge noise peaks as shown in Fig.~\ref{fig3},
to observe the sample noise level it is required that the spectrum
leakage and sidelobe background of the unwanted power line noise
peaks shouldn't mask the real white noise floor. This is usually
achieved by using special window function when doing FFT and by
increasing the frequency resolution\cite{Man00}. Since Hann
window has a fast decreasing sidelobe magnitude, it is preferred
in this situation than uniform window which is conventionally
used for flat noise spectrum measurement. And by increasing the
frequency resolution the peaks' mainlobe can be narrowed and
their sidelobes can be attenuated. In our case the sampling rate
is 44.1kHz, sampling number is chosen to be 32768 ($2^{15}$)
points for each step, so the sampling time $\Delta T$ is
$32768/44100=0.743$ second for each step, and frequency
resolution $\Delta f$ is $1/0.743=1.346$Hz. It is possible to
increase the sampling number to increase $\Delta T$ and decrease
$\Delta f$. This will require only larger PC memory and higher
speed CPU which is inexpensive. A simple algorithm is used here to
eliminate 3 points from both sides of those power line peaks
frequency when doing the average. This is already good enough to
find the real noise floor of curve B in Fig.~\ref{fig3}. More
complex ways using adaptive filter program to remove the noise
peak and extract the floor level is also possible.

\begin{figure}
\includegraphics{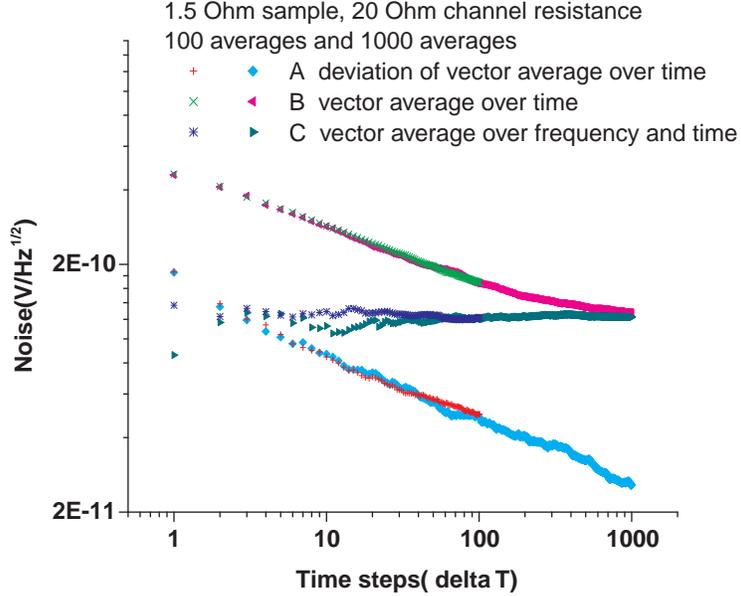}
\caption{\label{fig4} Two Cross-correlation test results for 1.5
$\Omega$ sample with 20$\Omega$ channel resistor, one stopped
after 100 time steps, the other stopped after 1000 time steps.
Curve B shows that using conventional average over time methods,
it approaches convergence limit after 1000 averages. Curve C
shows that using vector average over both frequency and time, the
convergence was approached within a few time steps.}
\end{figure}

As shown in Sec. \ref{cross}, the number of points used for
vector average over frequency domain decides how much times faster
of this new algorithm compared to conventional algorithm. Here
since we used the range from 1kHz to 2kHz, with resolution
1.364Hz, we get 733 points. After subtracting the number of those
points that are too close to noise peaks, we have around 600
points. So in principle we should get 600 times faster. To test
this we measured room temperature noise for 1.5$\Omega$ sample
resistor with same 20$\Omega$ channel resistors. The result is
shown in Fig.~\ref{fig4}. The start point of curve A and B are
mostly determined by the 20$\Omega$ resistor. There is a ratio
about 3 between those two curves around the start point. To
detect the noise level from 1.5$\Omega$ sample resistor,  we can
assume the required standard deviation $\sigma$ to be 5 times
smaller than the convergence limit $A$, which is
$\sqrt{4k_{B}TR_{1.5\Omega}}$, we would need $\sigma$ decrease to
magnitude $\sqrt{4k_{B}TR_{1.5\Omega}}/5$. The time needed for
conventional cross-correlation methods can be estimated by:
\begin{eqnarray}\label{eqn4}
(\frac{\sqrt{4k_{B}TR_{20\Omega}}/3}{\sqrt{4k_{B}TR_{1.5\Omega}}/5})^{4}\approx
1372
\end{eqnarray}
For the new algorithm, we expect $1372/600\approx 2$ steps.

As shown in Fig.~\ref{fig4} the convergence limit is about 0.126
nV/$\sqrt{Hz}$. It is close to the estimated value of thermal
noise level of a 1.5$\Omega$ resistor at room temperature, which
is 0.158 nV/$\sqrt{Hz}$. And $0.126/0.158\approx 0.8$ is
consistent with the 10$\Omega$ case. At 1000th time step, the
last point of curve A in Fig.~\ref{fig4} has the value 25.6
pV/$\sqrt{Hz}$. The ratio between convergence limit and standard
deviation is $126/25.6=4.9$, which is close to our estimation that
is 5 for 1372 steps. As for curve C, it approaches the
convergence limit from the first point in the 100 time step case,
in the 1000 time steps case,  despite of some fluctuations that
may caused by some broad band noise or data processing, it also
approaches the convergence limit from the first a few points. So
it is proved that with the algorithm of vector average over
frequency and time, convergence limit can be approached hundreds
of times faster than conventional cross-correlation algorithm with
this simple setup. If there are less noise peaks and if large
frequency band is available, this algorithm could give even
faster result.

\section{Conclusion}

For white noise measurement, an improved cross-correlation
algorithm using vector average over both frequency and time
domain is presented. With consideration of low temperature noise
measurement, a simple test set-up using PC and sound card is
built and tested. It is proved that this algorithm can achieve
convergence hundreds of times faster than the conventional
cross-correlation algorithm. Even with much bigger channel noise
and huge pick up noises, 100 pV/$\sqrt{Hz}$ noise level and
25pV/$\sqrt{Hz}$ sensitivity can be achieved in seconds. With a
broader frequency band width, better A/D card and larger PC memory
the convergence can be reached even faster. In principle this
algorithm could be used for other type of noises as long as the
shape of the spectrum is known and phase in frequency domain is
random(for example 1/f noise).

\end{document}